\documentclass[aps,pre,reprint,unsortedaddress,superscriptaddress]{revtex4-1} 

\usepackage{amsmath}  
\usepackage{amsfonts} 
\usepackage{graphicx} 

\usepackage{amssymb}

\usepackage{dcolumn}
\usepackage{bm}

\usepackage{verbatim}

\usepackage{color}
\usepackage{epstopdf}
\usepackage{hyperref}
\usepackage{xifthen}
\usepackage{changes}
\usepackage{ulem}

\DeclareMathAlphabet{\mathpzc}{OT1}{pzc}{m}{it}

\newlength\dlf  

\begin{document}



\title{Optimal control of aging in complex networks}

\author{Eric D. Sun}
\affiliation{Engineering and Applied Sciences, Harvard University, Cambridge, Massachusetts 02138, United States of America}
\affiliation{contributed equally}

\author{Thomas C. T. Michaels}
\affiliation{Engineering and Applied Sciences, Harvard University, Cambridge, Massachusetts 02138, United States of America}
\affiliation{contributed equally}

\author{L. Mahadevan}
\affiliation{
Engineering and Applied Sciences, Physics and Organismic and Evolutionary Biology, Harvard University, Cambridge, Massachusetts 02138, United States of America}

\begin{abstract}
Many complex systems experience damage accumulation which leads to aging, manifest as an increasing probability of system collapse with time. This naturally raises the question of how to maximize health and longevity in an aging system at minimal cost of maintenance and intervention. Here, we pose this question in the context of a simple interdependent network model of aging in complex systems, and use both optimal control theory and reinforcement learning alongside a combination of analysis and simulation to determine optimal maintenance protocols. These protocols may motivate the rational design of strategies for promoting longevity in aging complex systems with potential applications in therapeutic schedules and engineered system maintenance.
\end{abstract}



\maketitle 


\cleardoublepage

Aging is the process of damage accumulation with time that is responsible for an increasing susceptibility to death or decay \cite{harman_aging_1981}. Many complex systems that consist of multiple interacting components \cite{bar-yam_dynamics_2003}, e.g. biological organisms and artificially engineered systems, experience aging. Indeed, models of the interdependence between components of a system implemented in a network \cite{barabasi_network_2004} show aspects of aging and eventual system-wide catastrophe and death. This is because when components are interdependent, the failure of one component may adversely affect its dependents and vice versa. The dynamics of these processes have been the focus of many recent studies \cite{vural_aging_2014,taneja_dynamical_2016, farrell_network_2016,mitnitski_aging_2017}, exhibit temporal scaling \cite{stroustrup_temporal_2016,stroustrup_measuring_2018} and failure cascades, and reproduce experimental survivorship curves for many biological organisms and technological devices \cite{vural_aging_2014}. 

Understanding the onset of aging in network models point towards a central question in the field: how can one control aging in complex systems through interventions associated with repair and maintenance, with the eventual goal of designing strategies for increasing longevity \cite{lopez-otin_hallmarks_2013}? Available control strategies in networks are primarily for single nodes \cite{gao_target_2014} and sets of driver nodes \cite{liu_controllability_2011, cowan_nodal_2012}, and largely fall into three classes: network design \cite{zhao_tolerance_2005, tanizawa_optimization_2005}, edge and node removal at onset of cascade \cite{motter_cascade_2004}, and time-dependent edge weight distribution \cite{wang_universal_2008, mirzasoleiman_cascaded_2011}. Complementing these approaches, in reliability engineering there are maintenance policies for deteriorating multi-unit systems \cite{cho_survey_1991, thomas_survey_1986, pierskalla_survey_1976, wang_survey_2002} that include opportunistic repair \cite{radner_opportunistic_1963} and group and block replacement \cite{barlow_mathematical_1996} for systems with economic and structural dependencies between components \cite{thomas_survey_1986, pierskalla_survey_1976, sherif_optimal_1981}.  However, aging systems are primarily characterized by failure dependencies between components.  Only very special repair policies have been optimized for failure-dependent complex systems \cite{ross_model_1984}, and most are restricted to systems composed of few units \cite{thomas_survey_1986, pierskalla_survey_1976}, or with strong assumptions about the underlying failure distribution without consideration for the dynamical rules of individual network components from which they emerge \cite{wang_survey_2002, pham_imperfect_1996}.

Here we introduce a framework for determining optimal control strategies to delay aging in complex systems, modeled as interdependent networks, using two approaches, optimal control theory \cite{hocking_optimal_1991}, and  reinforcement learning \cite{watkins_q-learning_1992}, to derive explicit temporal repair protocols that control the healthspan and longevity of an interdependent aging system at minimum cost of intervention. Optimal control theory allows us to obtain mathematical expressions that characterize the optimal repair policy, while reinforcement learning provides a partially model-free approach by which the system may learn the optimal repair protocol.  Our work yields optimal repair policies   derived explicitly from a consideration of how the ``macroscopic" behavior of the network (decreasing vitality with failure cascades, see Fig.~\ref{fig1}b) emerges from the ``microscopic'' dynamics of the individual network components, which are interdependent and undergo stochastic failure and repair (Fig.~\ref{fig1}a), complementing previous approaches in network control \cite{crucitti_model_2004, motter_cascade_2004, zhao_tolerance_2005,liu_controllability_2011,gao_target_2014} and maintenance policies \cite{cho_survey_1991}, that do not discuss this micro-macro connection. Making this connection allows us to relate the optimal protocols to the underlying microscopic parameters, with natural quantitative interpretations in specific  systems.   Furthermore, in the context of reinforcement learning as an iterative updating process, the optimal maintenance policies that arise are similar to those in an optimal control context and suggest how iterative processes in evolution and biological learning may similarly arrive at these policies.

\section*{Network model of aging and repair}

\subsection*{Computational model} 
\begin{figure}[h!]
	\center\includegraphics[width=0.5\textwidth]{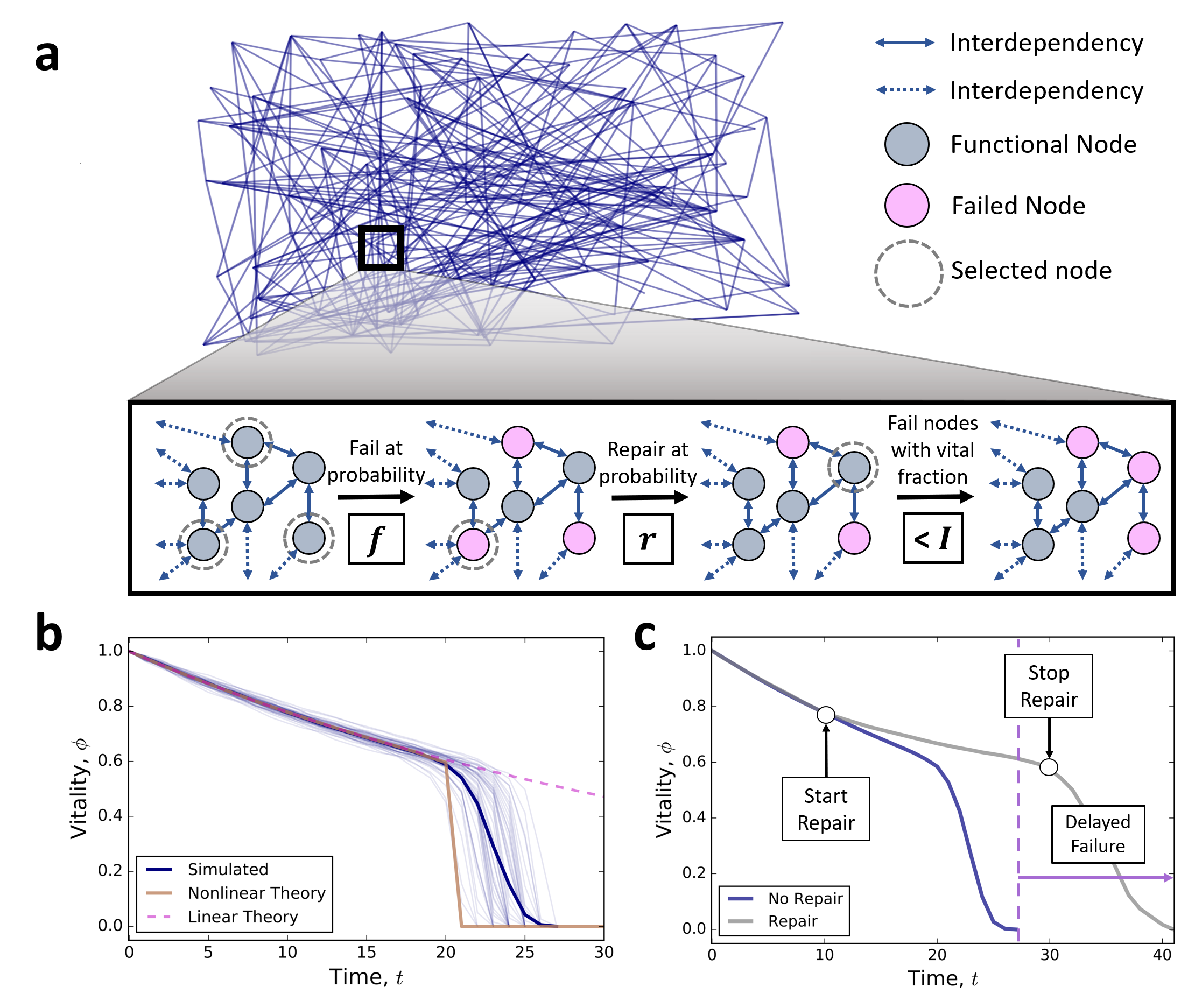}
	\caption{{\bf{Computational network model of aging and optimal repair.}} (a) Schematic representation of the network model of aging, represented by a network, where nodes denote components and edges denote interdependencies between these components. The network aging algorithm is portrayed in a smaller subsection of the network. At each time step, nodes are failed with probability $f$, repaired with probability $r$, and failed if their fraction of vital providers is less than $I$. (b) Simulated cascading failures in a Gilbert random model ($p\! =\!0.1$, $N\! =\!1000$, $f\! =\!0.025$, $r\! =\!0$, $d\! =\!0$, $I\! =\!0.5$). Faint blue lines refer to individual vitality trajectories, the solid blue line is the mean vitality $\Phi(t)$, the dashed magenta line is the analytic solution to the linear model, \eqref{eqphi}, where $I\! =\!0$, and the solid gold line is the numeric solution to the nonlinear model, \eqref{Eqphi}. (c) Network repair at $r\! =\!0.025$ from $T_1\! =\!10$ to $T_2\! =\!40$ (gray) delays network failure and improves lifespan and vitality as compared to a network without repair (blue).} 
	\label{fig1}
\end{figure}

Our computational model of aging starts with the consideration of a network with $N$ nodes representing the individual components of the complex system and edges between nodes representing interdependencies between the individual components (Fig.~\ref{fig1}a). The main network structure used in this study is the Gilbert $G(N,p)$ random graph \cite{gilbert_random_1959}; in this network, edges between any two nodes occur with probability $p$, where the mean node degree is $z=pN$. We also explore Erdos-Renyi $G(N,m)$ random networks \cite{erdos_evolution_1960} and Barabasi-Albert scale-free networks \cite{barabasi_emergence_1999}; these structures produce qualitatively similar results as compared to the Gilbert random graph (see Fig.~S4). In the model, each node is assigned an initial state of binary value $x_i \in\{0,1\}$ with probabilities $P(x_i\! =\!0) \! =\! d$ and $P(x_i\! =\!1)\! =\!1-d$, where $d$ denotes the prenatal damage of the complex system at birth. The state of a node represents its functionality, where $x_i \! =\! 1$ denotes a vital, functional $i$-th node and $x_i \! =\! 0$ denotes a dead, failed $i$-th node.

The network is then allowed to age via a simple iterative algorithm (see SI Algorithm S1) through the following actions: 1)   each node fails with probability $f$; 2)   nodes are repaired with probability $r$; 3) a node fails if the fraction of vital providers (i.e.~functional neighboring nodes) is less than $I$; 4)  the network vitality is calculated using the expression $\phi(t) \! =\! \frac{1}{N}\sum_{i=1}^N{x_i}$; 5)  the system fails if $\phi(t) < 0.1$ (Fig.~\ref{fig1}a).  Here, $I$ is a measure of the interdependence between the system components, and denotes the threshold fraction of vital providers required for a node to stay alive. Then $I\! =\!0$ corresponds to a collection of $N$ independent components, and if the vital fraction is less than $I$, then the node automatically fails. 

Our model reproduces the characteristic cascading failures that are present in the breakdown of complex systems \cite{crucitti_model_2004, motter_cascade_2004, zhao_tolerance_2005, vural_aging_2014}. In a representative simulation, the vitality $\phi(t)$ of the system decreases slowly in the linear regime before collapsing rapidly after a critical vitality value, $\phi_c$ (Fig.~\ref{fig1}b). The cascading failure is observed in all three graph structures (see Fig.~S4). This sudden decrease in system vitality is similar to the compression of morbidity that is observed during late life for humans and many other biological organisms \cite{fries_aging_2002}.

\subsection*{Nonlinear theory of network aging} 

To complement our computational model of aging networks, we also construct an effective equation for the average network vitality measured over several realizations, $\Phi(t) \! =\! \langle \phi(t) \rangle$. A mean-field model for the average vitality  as a function of time  may then be written as
\begin{equation}
\frac{d\Phi}{dt} = -f_{\rm{tot}} \Phi +r_{\rm{tot}} (1-\Phi),
\end{equation}
where $f_{\rm{tot}}\Phi$ is the total rate of node failure and $r_{\rm{tot}} (1-\Phi)$ is the total rate of repair. It is important to note that $f_{\rm{tot}}$ and $r_{\rm{tot}}$ denote the collective aspects of the network and are thus different from the respective intrinsic failure and repair rates $f$ and $r$ of nodes. They thus account for interdependence between nodes. To understand the relation between these variables, we note that a node fails for one of two reasons: 1) it fails with intrinsic rate $f$, or 2) it fails if the fraction of its vital provides falls below $I$ (i.e.~failure cascade). At leading order in failure rate $f$, we can neglect the simultaneous failure of two or more nodes at any time point; hence, induced failure occurs in one step, when the node is left with the minimum number of vital providers and then one of these vital providers fails. The total rate of node failure is thus given by the sum of the intrinsic failure rate $f\Phi$ and the rate of failure of the last vital provider
\begin{equation}\label{selfcons}
f_{\rm{tot}}\Phi = f \Phi + k  (1-f)\, f_{\rm{tot}} \, m(I,\Phi) ,
\end{equation}
where $k \! =\! zI $ is the minimum number vital providers required for a node to function, $z=pN$ is the average number of edges between nodes in the network (for a Gilbert random graph), and 
\begin{align}
m(I,\Phi) = \begin{pmatrix} z\\ k\end{pmatrix} \Phi^{k} (1-\Phi)^{z-k}
\end{align}
describes the (mean-field) probability that a node is left with $k$ vital providers. From \eqref{selfcons}, we obtain the total rate of failure as
\begin{equation}\label{m}
f_{\rm{tot}} = \frac{f}{1-k\frac{m(\Phi)}{\Phi}(1-f)}.
\end{equation}
 Similar arguments can be employed to determine the total rate of repair. A node can be repaired only if the following two conditions are met: 1) the node is failed, and 2) the node is connected to at least the minimum fraction $I$ of vital providers required for it to function once repaired. The total rate of repair is thus the product of the intrinsic rate of repair, $r(1-\Phi)$, and the probability $h(I,\Phi)$ that the node is connected to at least $k\! =\! zI$ vital providers: 
\begin{align}\label{h}
h(I,\Phi) = \sum_{j=k}^z  \begin{pmatrix} z\\ j\end{pmatrix} \Phi^j (1-\Phi)^{z-j}.
\end{align}
Combining \eqref{m} and \eqref{h}, we arrive at:
 \begin{equation}\label{Eqphi}
\frac{d\Phi}{dt} = -\frac{f\Phi}{1-k\frac{m(I,\Phi)}{\Phi}(1-f)} + r\,  h(I,\Phi) (1-\Phi),
\end{equation}
where $f$ and $r$ are the intrinsic frequencies of failure and repair, respectively, and interdependence between nodes is captured in this mean-field equation by the non-linear functions $m(I,\Phi)$ and $h(I,\Phi)$. In SI Fig.~S1, we compare the similarities and differences between the mean-field model \eqref{Eqphi} and the network simulations ( Fig.~\ref{fig1}b). Analytically, we see that the solution to \eqref{Eqphi} describes an average vitality that decreases slowly at early-times. In the limit when the system is away from collapse ($ft\ll 1$), \eqref{Eqphi} can be linearized and approximated to leading order as 
\begin{equation}\label{eqphi}
\frac{d\Phi}{dt} = -f \Phi +r (1-\Phi).
\end{equation} 
This leads to an exponentially decaying vitality (Fig.~\ref{fig1}b). At later times, the average vitality exhibits failure cascade and rapid collapse after a critical vitality value $\Phi_c$ is reached (Fig.~\ref{fig1}b). This effect originates when the denominator in the first term on the right-hand side of \eqref{Eqphi} becomes small, which causes the effective failure rate to blow up; thus, an estimate for the critical fraction for failure cascade can be obtained by
maximizing the function $m(\Phi)/\Phi$ over $\Phi$, which yields for large $z\gg 1$ (see SI Sec.~S1):
\begin{equation}
\Phi_c \simeq I.
\end{equation}
Note that \eqref{Eqphi} is similar to a model previously proposed in Ref.~\cite{vural_aging_2014}; however, our \eqref{Eqphi} exhibits cascading failures, while the model in Ref.~\cite{vural_aging_2014} does not.

\section*{Optimal control of network aging}

Having a qualitative understanding of the forward problem of understanding how aging arises in interdependent networks, we now turn to the problem of controlling the progressive aging of a network by varying the repair rate, subject to some constraints. 

\subsection*{Optimal repair protocols}

For an interdependent network that ages according to \eqref{Eqphi}, our goal is to design optimal repair protocols, i.e. replace the constant repair frequency $r$ in \eqref{Eqphi} by a time dependent unknown repair rate $r(t)$ to regulate network vitality. Since high vitality is expected to correspond to a ``benefit'', while repair actions come with a ``cost'', we introduce the following cost function to capture this balance between network vitality and repair:
\begin{equation}\label{2} 
\text{Cost} = \int_{0}^{T} e^{-\gamma t}\, \mathcal{C} (\phi(t),r(t)) dt,
\end{equation}
where $T$ is the final time and $\mathcal{C} $ is a monotonically decreasing function of vitality $\phi(t)$ and a monotonically increasing function of repair $r(t)$.
The exponential term describes the situation when future values of the cost are discounted, where $\gamma \geq 0$ is the discount rate. We focus here on a simple linear cost function $ \mathcal{C} \! =\! \alpha r(t) - \phi(t),$ where $\alpha$ is the relative cost of repair, but note that our approach can be generalized to arbitrary cost functions (see SI for details). The first term in the linear cost function describes the total cost for repair as the integral of all repair events that have occurred in time, while the second term is the gain from vitality; the constant $\alpha$ describes the relative importance of the two terms in the cost function. The goal of the optimal control problem defined by \eqref{Eqphi} and \eqref{2} is to find the repair protocol $r(t)$ that minimizes the cost function \eqref{2} while satisfying the evolution equation \eqref{Eqphi} for vitality. 

We solve this optimal control problem for a network with initial vitality $\Phi(t\! =\!0)\! =\!1-d$ using the framework of optimal control theory and Pontryagin's  principle \cite{hocking_optimal_1991} (see SI Sec.~S2 for details). Since the optimal control problem is linear in the repair rate $r(t)$, the optimal repair protocol will correspond to a bang-bang control that switches between $r(t)\! =\!0$ (no repair) and $r(t)\! =\!r$ (maximal repair); repair is turned on when the following condition is met:
\begin{equation}\label{cond}
h(I,\Phi)(1-\Phi) \geq \frac{\alpha}{|\lambda|},
\end{equation}
where $\lambda$ is a time-dependent co-state variable, which is determined as the solution to Eq.~S22 (see SI Sec.~S2). \eqref{cond} admits an interesting physical interpretation. It states that the optimal decision to repair depends on two parameters: 1) the repairable fraction of nodes, $h(I,\Phi)(1-\Phi)$, and 2) a time-dependent threshold $\alpha/|\lambda|$, which depends on the relative cost of repair $\alpha$. The repairable fraction increases with time as nodes in the network fail and/or become increasingly susceptible to failure cascades; on the other hand, the threshold for the repairable fraction also increases with time as the system ages, leading to a smaller window of repair. These two opposing effects lead to non-monotonic optimal repair protocols characterized by a waiting time for repair, followed by an intermediate period where repair is preferable and a terminal phase where the repair rate is set again to zero (see SI Fig.~S2). Mathematically:
\begin{equation}\label{Tsbb}
r(t) = \begin{cases}
0, & t< T_1 \\
r, & T_1\leq t< T_2 \\
0, & t\geq T_2
\end{cases}
\end{equation}
where $T_1$ and $T_2$ are switching times.

\subsection*{Linear control theory}

To gain an understanding of how the optimal repair protocol depends on the physical parameters, it is useful to focus first on the linearized limit, corresponding to \eqref{eqphi}, which is valid when the system is away from  vitality collapse. In fact, explicit analytical expressions for the switching times can be obtained in this case (see SI Sec.~S2 for a derivation):
\begin{subequations}\label{Ts}
\begin{align}
T_1 & \simeq \frac{1}{f}\log\left[\frac{1-d}{1-\alpha(f+r+\gamma)}\right]\label{Tsa}\\
T_2 & \simeq T-\frac{1}{f+\gamma}\log\left[\frac{1}{1-\alpha(f+r+\gamma)(f+\gamma)/f}\right].\label{Tsb}
\end{align}
\end{subequations}
The dependence of $T_1$ and $T_2$ on the failure rate $f$, repair rate $r$, and cost of repair $\alpha$ is shown in Fig.~\ref{fig2}. The optimal repair protocol in time consists of an initial phase when system vitality is high and no repair is necessary and a repair period that is initiated at time $T_1$ and persists until time $T_2$.
For $\gamma\! =\!0$ and $d\! =\!0$ (corresponding to a healthy organism), the repair protocol is symmetric with respect to the end time $T$, since $T_1\! =\!T-T_2$. The protocol is no longer symmetric with respect to $T$ when $d>0$; in particular, while the initial vitality level does not affect the end time $T_2$, the start time $T_1$ decreases with increasing $d$, implying that the optimal repair protocol starts earlier and lasts for longer as the initial vitality of the system decreases. There is a critical value for initial vitality, $\Phi(t\! =\!0) < 1-d_c \! =\! 1-\alpha(f+r+\gamma)$, below which the optimal repair protocol starts right away. 

In the infinite horizon limit $T \rightarrow \infty$ and $\gamma > 0$, we enter a regime where the optimal solution for repair maximizes the discounted health of the system over an indefinite period of time under a cost constraint. Biologically, this is equivalent to optimizing longevity as compared to healthspan for finite $T$, while considering a discount factor resulting from extrinsic mortality \cite{kirkwood_understanding_2005}. Since $T_2 \rightarrow \infty$, the infinite horizon repair protocol is characterized by a single switching time $T_1$, after which the system is repaired in perpetuity. 

Thus far, we have focused on the simple linear cost function. Exploring non-linear cost functions leads to the optimal repair protocol that is no longer bang-bang, but is still non-monotonic in time (SI Sec.~S6), with initial and terminal phases of low repair and an intermediate region of higher repair (see Fig.~S5). Additoinal extensions may be motivated by future experiments and might involve considering a terminal cost for vitality, including nonlinearities in vitality and/or repair rate (see SI Sec.~S6) or introducing additional variables, such as node checking and associated cost (see SI Sec.~S7).

\subsection*{Phase diagram for repair}
\begin{figure}[h!]
\center\includegraphics[width=0.45\textwidth]{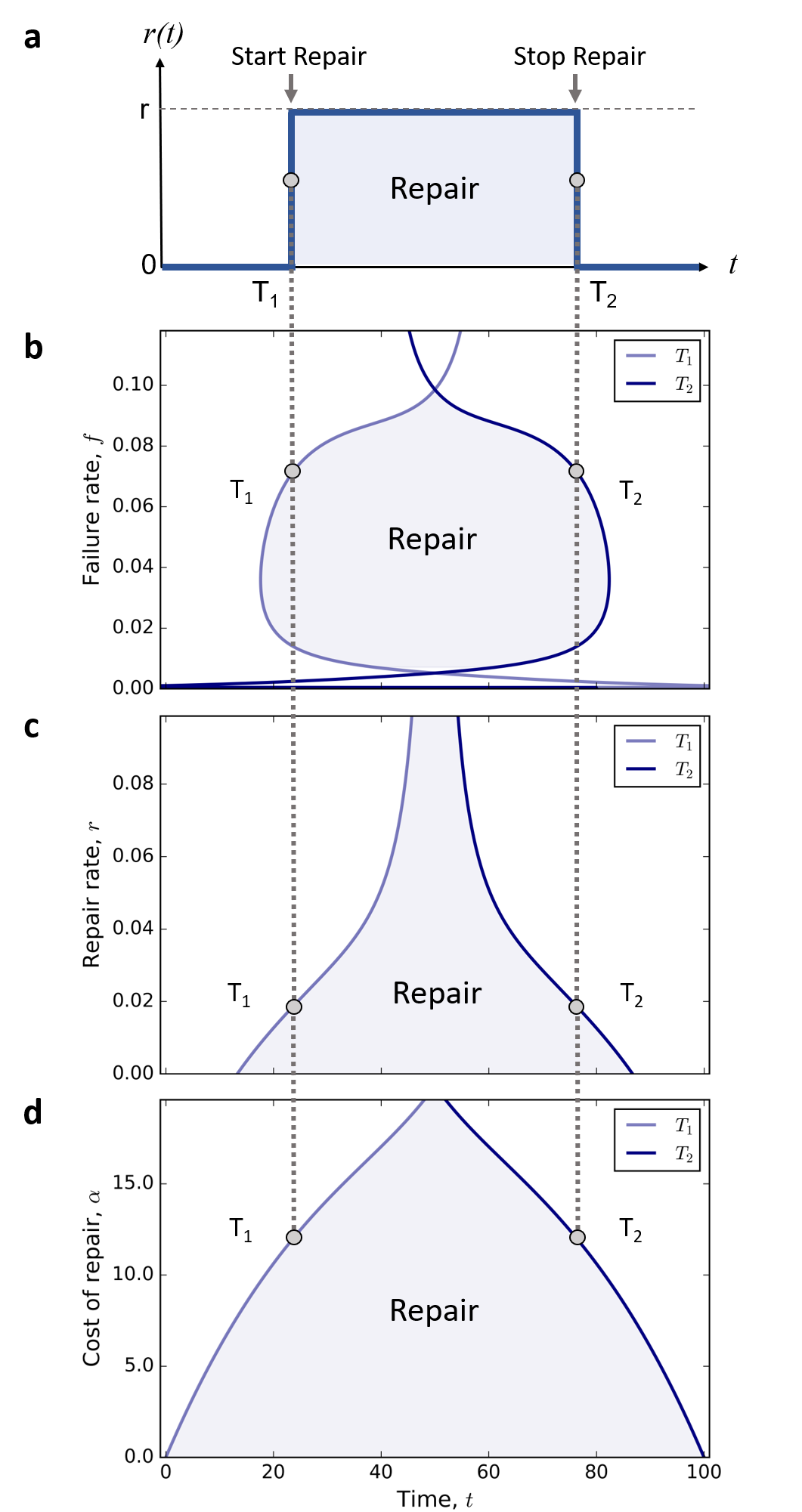}
\caption{{\bf{Optimal repair protocols to maximize healthspan at minimum intervention cost.}} (a) Schematic representation of optimal bang-bang repair protocol $r(t)$ with repair start time $T_1$ and repair stop time $T_2$ as showcased in \eqref{Tsbb} for the linear regime. (b) The repair duration (shaded blue) is dependent on the failure rate $f$ and disappears for small $f$ and large $f$ as calculated from \eqref{Ts}. (c) The repair duration monotonically decreases with increased maximum repair rate $r$. (d) The repair duration decreases with increased cost of repair $\alpha$ and disappears for large $\alpha$. The default parameters used were $N\! =\!1000$, $p\! =\!0.1$, $f\! =\!0.025$, $r\! =\!0.01$, $\alpha\! =\!10$, $\gamma\! =\!0$, $T\! =\!100$, $d\! =\!0$, $I\! =\!0$.
}
\label{fig2}
\end{figure}

A question of some interest is the determination of  the conditions under which a repair protocol is advisable. From  \eqref{Ts}, it follows that since $T_1$ must, by definition, be smaller than $T_2$, a repair protocol exists for $d\! =\!0$ and $\gamma\! =\!0$ only if: 
\begin{equation}\label{phasediag}
fT \geq  2\log\left[\frac{1}{1-\alpha(f+r)}\right] .
\end{equation}
\eqref{phasediag} results in a phase diagram separating a region of ``repair'' from a region of ``no repair'', where repair is too costly, as a function of two relevant dimensionless parameters $\alpha(f+r)$ and $fT$. As a function of failure frequency $f$ and at constant values of $\alpha,r$ and $T$, \eqref{phasediag} predicts the existence of regions of low ($fT\ll 1$), and respectively, high failure rates ($fT\gg 1$), where the best option is not to repair (Fig.~\ref{fig2}b). This behavior follows intuition; when failure rate is low, vitality remains high over the interval $[0,T]$, such that the cost of repair would be unnecessarily large compared to the benefit associated with increased vitality. Similarly, when the failure rate is large a significant improvement of vitality would require an insurmountable cost of repair. As the repair rate $r$ increases, \eqref{phasediag} predicts a rapidly shrinking window of repair due to the combined effect of increasing the effectiveness of and associated cost ($\alpha r$) of repair (Fig.~\ref{fig2}c). As the cost of repair $\alpha$ increases, \eqref{phasediag} similarly predicts a decreasing window of repair (Fig.~\ref{fig2}d) that results from an increasing cost burden. There exists a critical value for the repair cost, $\alpha_c \! =\! 1/(f+r)$, above which there is no repair.

\subsection*{Interdependent networks} For networks with interdependent components, the optimal protocols are still bang-bang and the switching times can be calculated using \eqref{cond}.

\begin{figure}[h!]
	\center\includegraphics[width=0.4\textwidth]{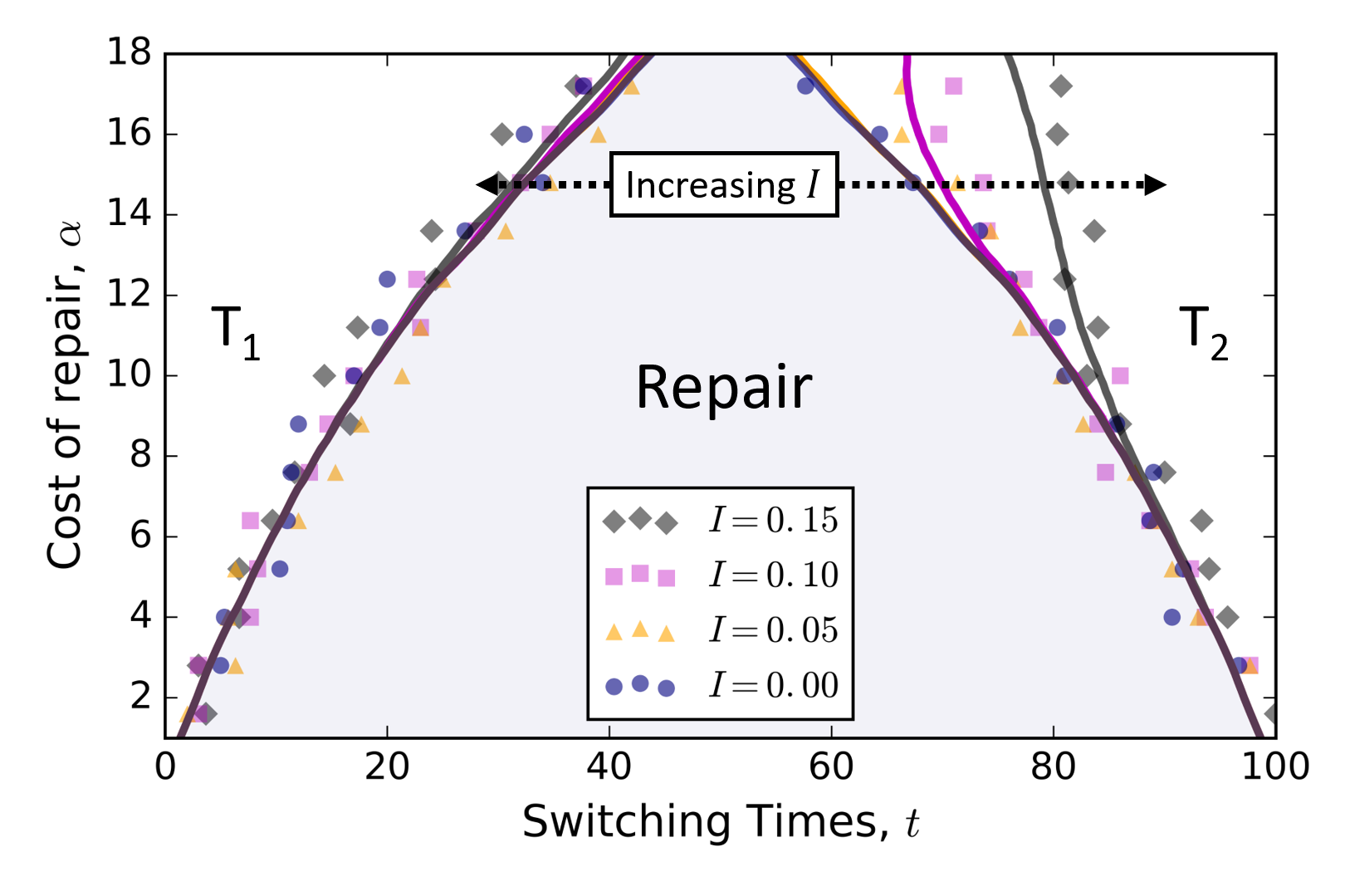}
	\caption{{\bf{Optimal repair protocol for an interdependent network.}} As $I$ increases, the stop time $T_2$ for the repair protocol increases while the start time $T_1$ decreases marginally. Solid lines correspond to the numerical solution to the optimal control problem. Scatter points correspond to the optimal switching times obtained from a grid search on the computational model. The default parameters used were $N\! =\!1000$, $p\! =\!0.1$, $f\! =\!0.025$, $r\! =\!0.01$, $\gamma\! =\!0$, $T\! =\!100$, $d\! =\!0$.
	}
	\label{fig22}
\end{figure}

Notably, increasing the interdependence ($I \geq 0$) between components provided qualitatively similar strategies for maintaining optimal healthspan (finite $T$) as the linear theory. Our theory predicts that the window of repair increases with interdependence in order to compensate for the accelerated aging and reduced response to repair in interdependent networks. Increasing $I$ has little effect on the switching time $T_1$, since at high vitality the interdependent system is close to the linear theory. However, as $I$ increases, the repairable $h(\Phi, I)$ fraction and the effective repair rate decrease monotonically with $I$ for fixed $\Phi$, which results in an increasing repair stop time $T_2$. We ran computational simulations of the network model to validate the predicted optimal repair policies as interdependence is increased (see SI Sec.~S3 for details on the simulations). The results shown in Fig.~\ref{fig22} agree with the optimal policies calculated using \eqref{cond} (solid lines).

\begin{figure*}
	\center\includegraphics[width=0.96\textwidth]{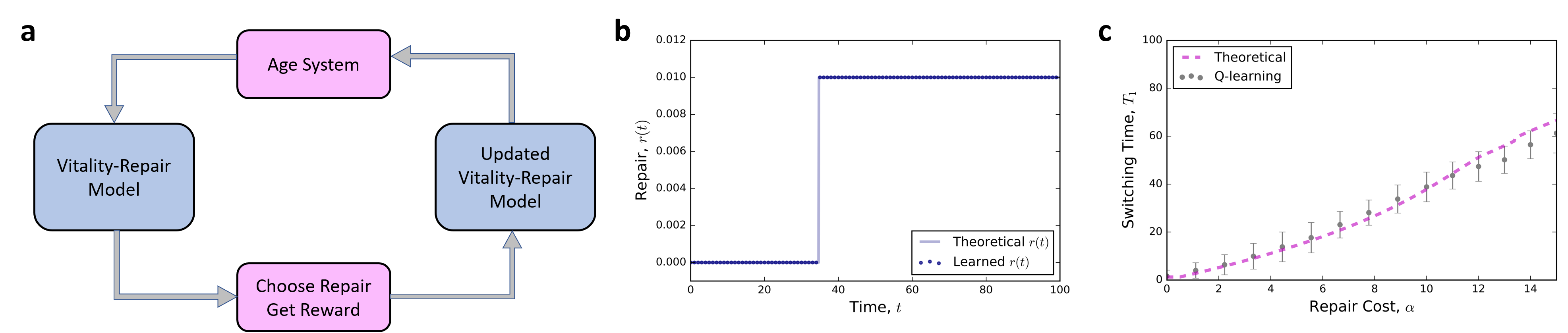}
	\caption{{\bf{Optimal repair protocols using reinforcement learning.}} (a) High-level schematic of reinforcement learning algorithm for optimal control of network aging. Refer to SI Sec.~S4 for further Q-learning model details. (b) The learned repair protocol (represented as points) is bang-bang and matches closely with the theoretically optimal repair protocol (line, see \eqref{Tsbb}) and is characterized by a single repair switching time $T_1$. Parameters used were $f\! =\!0.0367$, $r\! =\!0.01$, $\alpha\! =\!10$, $\gamma\! =\!0.975$, $I\! =\!0$, $d\! =\!0$. (c) Optimal $T_1$ as a function of the cost of repair $\alpha$ for the reinforcement learning (gray circles), $N\! =\!50$ realizations) and the theoretical solution (dotted magenta, see \eqref{Ts}). Models used $N\! =\!1000$, $p\! =\!0.1$, $f\! =\!0.025$, $r\! =\!0.01$, $\gamma\! =\!0.975$, $I\! =\!0$.}
	\label{fig3}
\end{figure*}

\subsection*{Role of network topology}
Thus far, we have presented optimal controls for complex systems with random structures \cite{gilbert_random_1959}. We have also studied optimal protocols numerically for Erdos-Renyi $G(N,m)$ random networks \cite{erdos_evolution_1960} and Barabasi-Albert scale-free networks \cite{barabasi_emergence_1999}. The aging dynamics are highly similar between the three network models investigated (see SI Fig.~S4a-c). For all random and scale-free networks, we observe no significant qualitative differences in the optimal repair protocols (see SI Fig.~S4d,e), indicating that our protocols are robust and may be applicable to a diverse range of complex systems.

\section*{Reinforcement learning approach to interdependent network aging control} 

Optimal control strategies rely on knowledge of both the model and a cost function, both of which are hard to crystalize into quantitative form in many biological systems. An alternate strategy is to ask  whether the system is able to learn the optimal repair protocol for aging via  an iterative procedure, tantamount to direct adaptive optimal control \cite{Sutton}. One possibility towards this end is the use of a machine learning strategy known as reinforcement learning, the process by which a system is able to optimize its actions by interacting with its environment. Optimization occurs iteratively on a trial and error basis, since every action corresponds to a reward/punishment. Through this process, optimal decisions that maximize reward and/or minimize punishment are reinforced. We use a relatively simple version of this algorithm known as the Q-learning model (Fig.~\ref{fig3}a, see Supplementary Fig.~S3) \cite{watkins_q-learning_1992}. This consists of creating a Q-matrix, $Q\! =\!$\{$\phi$, $r(\phi)$\}, which serves as a look-up table of vitality states $\phi$ and values associated with each possible action, $r(\phi)\! =\!0$ or $r(\phi)\! =\!r$. In each training episode, a healthy ($d\! =\!0$) network is initialized. At each time step, the network is subjected to the aging algorithm and the agent exploits network repair for the greatest-valued choice of repair at the given vitality of the system with probability $1-e^{-\lambda_{\text{exp}}q}$ where $q$ is the number of episodes elapsed. The agent explores with probability $e^{-\lambda_{\text{exp}}q}$. A reward $R$ is calculated and used to update the state-action value in the Q-matrix according to the rule \cite{watkins_q-learning_1992}
$$
    Q(\phi_t,r_t) \leftarrow Q(\phi_t,r_t) + \beta[R_{t+1} + \gamma_Q \max_r{Q(\phi_{t+1}}-Q(\phi_t,r_t)],
$$
$$
    R_t = \phi_t - \alpha r_t,
$$
where $\alpha$ is the cost of repair, $\beta$ is the learning rate, and $\gamma_Q$ is the Q-learning discount factor that is related to the optimal control through $\gamma \! =\! -\log \gamma_Q$. The learning rate exponentially decays as $\beta = e^{-\lambda_{\beta} q}$. An episode ends when the network fails (i.e.~$\phi < 0.1$). The Q-learning model iterates through learning episodes until qualitative convergence of the Q-matrix is achieved. The optimal protocol is defined as the maximal Q-valued trajectory traveled by a network through $(\phi, r)$ space. 

Using this method, the Q-learned repair policy converges at optimal repair protocols that are bang-bang (Fig.~\ref{fig3}b) and closely match the predicted switching time $T_1$ from the analytic theory for different values of $\alpha$ (Fig.~\ref{fig3}c). 
These results suggest that the optimal protocols for repair can be obtained through simple iterative learning and highlight the potential of Q-learning as a method to approximate optimal repair protocols for complicated systems in which no analytic description of the aging dynamics is available.

\section*{Discussion}

Although aging in real biological and technological systems clearly results from complex biochemical and mechanical processes, here we have abstracted a minimal model for  aging designed to capture the essential ingredients that give rise to aging in a complex system - modular units (nodes) that are interdependent via a set of edges modeled as an interdependent network subject to nodal failure and repair. Our model shows  the emergence of failure cascades, a hallmark of such systems.  Having understood how aging arises in this model, we showed how to derive optimal protocols for controlling aging in such systems.  First, we used a model dependent strategy, using optimal control theory  to determine explicit optimal repair protocols for aging interdependent systems characterized by a failure rate $f$, repair rate $r$, and interdependency $I$. We also demonstrated that a model-free approach, using reinforcement learning converges to these optimal repair protocols and can therefore be leveraged to approximate optimal repair strategies in an iterative manner, perhaps through evolution via natural selection. 

Our approach may motivate the design of treatments for maximizing healthspan and longevity in biological populations and/or prolonging the functionality of technological systems. For instance, the optimal repair protocols may potentially be applied to optimizing treatments targeting the clearance of senescent cells. Senescent cells are those that enter a permanent, non-dividing state and adopt an altered secretory profile referred to as SASP (senescence-associated secretory phenotype) that has been implicated in inflammation, tumorigenesis, and aging \cite{campisi_cellular_2007,lopez-otin_hallmarks_2013}. Furthermore, these senescent cells have been shown to promote the senescence of healthy cells in surrounding tissue \cite{xu_senolytics_2018}, which is similar to our model where node failure can spread due to interdependence. The selective clearance of senescent cells (i.e.~via the use of senolytic cocktails) improves physical function and survival in mouse models \cite{baker_clearance_2011,xu_senolytics_2018}.
These senolytic treatments do not significantly reduce the total cell count in human tissue nor do they decrease the body weight of mouse models \cite{xu_senolytics_2018}. This suggests a rapid replacement of cleared senescent cells by healthy dividing cells. In this limit, the application of senolytic treatments becomes analogous to node repair in an aging network. Moreover, many senolytic cocktails observe toxicity \cite{serrano_targeting_2018,kirkland_cellular_2017}, which mimics the cost of repair in our model. The relative cost $\alpha$ could be determined by separately measuring and then comparing the loss of vitality caused by senescence to the toxicity that results from senolytic cocktails on an ensemble of healthy cells. Possible measures of cellular vitality include the proportion of non-senescent cells and standard cell viability metrics. These optimal repair protocols may therefore motivate the design of treatment schedules for senescent cell inhibitors and other therapeutics that target general aging processes and/or extend healthspan. Natural next steps include generalizing our approaches to account for spatial organization and fluctuations. 

\textbf{Acknowledgements}\\
{We acknowledge support from the Swiss National Science Foundation (TCTM) and the Amgen Scholars Program (EDS).}



\begin{thebibliography}{10}

\bibitem{harman_aging_1981}
Harman D (1981) The aging process.
\newblock {\em Proceedings of the National Academy of Sciences}
  78(11):7124--7128.

\bibitem{bar-yam_dynamics_2003}
Bar-Yam Y (2003) {\em Dynamics of {Complex} {Systems}}.
\newblock (Westview Press).

\bibitem{barabasi_network_2004}
Barabási AL, Oltvai ZN (2004) Network biology: understanding the cell's
  functional organization.
\newblock {\em Nature Reviews Genetics} 5(2):101--113.

\bibitem{vural_aging_2014}
Vural DC, Morrison G, Mahadevan L (2014) Aging in complex interdependency
  networks.
\newblock {\em Physical Review E} 89(2):022811.

\bibitem{taneja_dynamical_2016}
Taneja S, Mitnitski AB, Rockwood K, Rutenberg AD (2016) Dynamical network model
  for age-related health deficits and mortality.
\newblock {\em Physical Review E} 93(2):022309.

\bibitem{farrell_network_2016}
Farrell SG, Mitnitski AB, Rockwood K, Rutenberg AD (2016) Network model of
  human aging: {Frailty} limits and information measures.
\newblock {\em Physical Review E} 94(5):052409.

\bibitem{mitnitski_aging_2017}
Mitnitski AB, Rutenberg AD, Farrell S, Rockwood K (2017) Aging, frailty and
  complex networks.
\newblock {\em Biogerontology} 18(4):433--446.

\bibitem{stroustrup_temporal_2016}
Stroustrup N, et~al. (2016) The temporal scaling of \textit{{Caenorhabditis}
  elegans} ageing.
\newblock {\em Nature} 530(7588):103--107.

\bibitem{stroustrup_measuring_2018}
Stroustrup N (2018) Measuring and modeling interventions in aging.
\newblock {\em Current Opinion in Cell Biology} 55:129--138.

\bibitem{lopez-otin_hallmarks_2013}
López-Otín C, Blasco MA, Partridge L, Serrano M, Kroemer G (2013) The
  {Hallmarks} of {Aging}.
\newblock {\em Cell} 153(6):1194--1217.

\bibitem{gao_target_2014}
Gao J, Liu YY, D'Souza RM, Barabási AL (2014) Target control of complex
  networks.
\newblock {\em Nature Communications} 5:5415.

\bibitem{liu_controllability_2011}
Liu YY, Slotine JJ, Barabási AL (2011) Controllability of complex networks.
\newblock {\em Nature} 473(7346):167--173.

\bibitem{cowan_nodal_2012}
Cowan NJ, Chastain EJ, Vilhena DA, Freudenberg JS, Bergstrom CT (2012) Nodal
  {Dynamics}, {Not} {Degree} {Distributions}, {Determine} the {Structural}
  {Controllability} of {Complex} {Networks}.
\newblock {\em PLOS ONE} 7(6):e38398.

\bibitem{zhao_tolerance_2005}
Zhao L, Park K, Lai YC, Ye N (2005) Tolerance of scale-free networks against
  attack-induced cascades.
\newblock {\em Physical Review E} 72(2):025104.

\bibitem{tanizawa_optimization_2005}
Tanizawa T, Paul G, Cohen R, Havlin S, Stanley HE (2005) Optimization of
  network robustness to waves of targeted and random attacks.
\newblock {\em Physical Review E} 71(4):047101.

\bibitem{motter_cascade_2004}
Motter AE (2004) Cascade {Control} and {Defense} in {Complex} {Networks}.
\newblock {\em Physical Review Letters} 93(9):098701.

\bibitem{wang_universal_2008}
Wang WX, Chen G (2008) Universal robustness characteristic of weighted networks
  against cascading failure.
\newblock {\em Physical Review E} 77(2):026101.

\bibitem{mirzasoleiman_cascaded_2011}
Mirzasoleiman B, Babaei M, Jalili M, Safari M (2011) Cascaded failures in
  weighted networks.
\newblock {\em Physical Review E} 84(4):046114.

\bibitem{cho_survey_1991}
Cho DI, Parlar M (1991) A survey of maintenance models for multi-unit systems.
\newblock {\em European Journal of Operational Research} 51(1):1--23.

\bibitem{thomas_survey_1986}
Thomas LC (1986) A survey of maintenance and replacement models for
  maintainability and reliability of multi-item systems.
\newblock {\em Reliability Engineering} 16(4):297--309.

\bibitem{pierskalla_survey_1976}
Pierskalla WP, Voelker JA (1976) A survey of maintenance models: {The} control
  and surveillance of deteriorating systems.
\newblock {\em Naval Research Logistics Quarterly} 23(3):353--388.

\bibitem{wang_survey_2002}
Wang H (2002) A survey of maintenance policies of deteriorating systems.
\newblock {\em European Journal of Operational Research} 139(3):469--489.

\bibitem{radner_opportunistic_1963}
Radner R, Jorgenson DW (1963) Opportunistic {Replacement} of a {Single} {Part}
  in the {Presence} of {Several} {Monitored} {Parts}.
\newblock {\em Management Science} 10(1):70--84.

\bibitem{barlow_mathematical_1996}
Barlow RE, Proschan F (1996) {\em Mathematical {Theory} of {Reliability}}.
\newblock (SIAM).
\newblock Google-Books-ID: wDDib1jBgtYC.

\bibitem{sherif_optimal_1981}
Sherif YS, Smith ML (1981) Optimal maintenance models for systems subject to
  failure–{A} {Review}.
\newblock {\em Naval Research Logistics Quarterly} 28(1):47--74.

\bibitem{ross_model_1984}
Ross SM (1984) A model in which component failure rates depend on the working
  set.
\newblock {\em Naval Research Logistics Quarterly} 31(2):297--300.

\bibitem{pham_imperfect_1996}
Pham H, Wang H (1996) Imperfect maintenance.
\newblock {\em European Journal of Operational Research} 94(3):425--438.

\bibitem{hocking_optimal_1991}
Hocking LM (1991) {\em Optimal {Control}: {An} {Introduction} to the {Theory}
  with {Applications}}.
\newblock (Clarendon Press).
\newblock Google-Books-ID: gd7b4FMqXpMC.

\bibitem{watkins_q-learning_1992}
Watkins CJCH, Dayan P (1992) Q-learning.
\newblock {\em Machine Learning} 8(3):279--292.

\bibitem{crucitti_model_2004}
Crucitti P, Latora V, Marchiori M (2004) Model for cascading failures in
  complex networks.
\newblock {\em Physical Review E} 69(4):045104.

\bibitem{gilbert_random_1959}
Gilbert EN (1959) Random {Graphs}.
\newblock {\em The Annals of Mathematical Statistics} 30(4):1141--1144.

\bibitem{erdos_evolution_1960}
Erdős P, Rényi A (1960) On the {Evolution} of {Random} {Graphs}.
\newblock {\em Publ. Math. Inst. Hung. Acad. Sci} p.~45.

\bibitem{barabasi_emergence_1999}
Barabási AL, Albert R (1999) Emergence of {Scaling} in {Random} {Networks}.
\newblock {\em Science} 286(5439):509--512.

\bibitem{fries_aging_2002}
Fries JF (2002) Aging, natural death, and the compression of morbidity.
\newblock {\em Bulletin of the World Health Organization} 80:245--250.

\bibitem{kirkwood_understanding_2005}
Kirkwood TBL (2005) Understanding the {Odd} {Science} of {Aging}.
\newblock {\em Cell} 120(4):437--447.

\bibitem{Sutton}
Sutton R, Barto A, Williams R (1992) Reinforcement learning is direct adaptive
  optimal control.
\newblock {\em IEEE Control Systems Magazine} 12:19--22.

\bibitem{campisi_cellular_2007}
Campisi J, d'Adda~di Fagagna F (2007) Cellular senescence: when bad things
  happen to good cells.
\newblock {\em Nature Reviews Molecular Cell Biology} 8(9):729--740.

\bibitem{xu_senolytics_2018}
Xu M, et~al. (2018) Senolytics improve physical function and increase lifespan
  in old age.
\newblock {\em Nature Medicine} 24(8):1246.

\bibitem{baker_clearance_2011}
Baker DJ, et~al. (2011) Clearance of p16$^{\textrm{{ink}4a}}$-positive
  senescent cells delays ageing-associated disorders.
\newblock {\em Nature} 479(7372):232--236.

\bibitem{serrano_targeting_2018}
Serrano M, Barzilai N (2018) Targeting senescence.
\newblock {\em Nature Medicine} 24(8):1092.

\bibitem{kirkland_cellular_2017}
Kirkland JL, Tchkonia T (2017) Cellular {Senescence}: {A} {Translational}
  {Perspective}.
\newblock {\em EBioMedicine} 21:21--28.


\end{thebibliography}
\end{document}